\documentclass{PoS}
\usepackage{collref}
\newcommand{\lb}{\langle \kern-.17em \langle} 
\newcommand{\rb}{\rangle \kern-.17em \rangle }

\title{A novel density of state method for complex action systems}

\ShortTitle{A novel density of state method for complex action systems}

\author{\speaker{Biagio Lucini}\\
         College of Science, Swansea University, Swansea SA2 8PP, UK\\
         E-mail: \email{b.lucini@swansea.ac.uk}}

\author{Kurt Langfeld\\
        School of Computing \& Mathematics, Plymouth, PL4 8AA, UK\\
        E-mail: \email{kurt.langfeld@plymouth.ac.uk}}

\abstract{Recently, a new and efficient algorithm (the LLR method) has been proposed for computing densities of states in statistical systems and gauge theories. In this talk, we explore whether this novel density of states method can be applied to numerical computations of observables in systems for which the action is complex. To this purpose, we introduce a generalised density of states, in terms of which integrals of oscillating observables can be determined semi-analytically, and we define a strategy to compute it with the LLR method. As a case study, we apply these ideas to the $\mathbb{Z}(3)$ spin model at finite density, finding a remarkable agreement of our results for the phase twist with those obtained with the worm algorithm for all explored chemical potentials, including values for which there are cancellations over sixteen orders of magnitude. These findings open new perspectives for dealing with the sign problem on physically more relevant systems.}

\FullConference{The 32nd International Symposium on Lattice Field Theory\\
   23-28 June, 2014\\
    Columbia University New York, NY}

\begin{document}

\section{Introduction and motivations}
Monte-Carlo simulations of the theory discretised on a spacetime
lattice provide a first principle method to compute observables in
QCD at zero baryon density. At the heart of this approach is the
interpretation of the Euclidean path integral measure as a Boltzmann
weight. This relies on the measure itself being positive. However, as
soon as a chemical potential coupled to the baryon number is switched
on, the measure becomes complex and importance sample methods are no
longer of use. In fact, because of the action being complex, the
path integral has contributions alternating in sign that give rise to
severe numerical cancellations. This phenomenon is called the sign
problem and characterises not only QCD at finite density, but dense
quantum systems in general. Other cases in which the sign problem
hinder numerical simulations with importance sampling methods include
interactions with external electromagnetic fields, rotating frames and
real-time dynamics.

For numerical simulations of those systems, radically different
approaches are needed. Significant progress has been achieved
recently, with the introduction of several methods based on a wide
variety of ideas (see~\cite{Aarts:2013bla,Gattringer:2014cpa} for
recent reviews). Most of these algorithms are still in their
infancy. In particular, both their numerical and mathematical
properties need to be better understood. For this reason, it is
particularly useful from an empirical point of view to compare them on
the same models, in order to understand and assess their strengths. In
this work, we give an overview of a recently proposed
method~\cite{Langfeld:2014nta} based on a novel
algorithm~\cite{Langfeld:2012ah} for the computation of the density of
states. The underlying idea is that if a suitable generalised density
of states is defined, the complex integral can be reduced to a
one-dimensional oscillatory integral. If sufficient precision is
available on the computation of the density of states, these types of
integrals can be done with the desired accuracy. We remark that the
general strategy is not new~\cite{Gocksch:1988iz} and has been revisited several
times (see e.g.~\cite{Azcoiti:2011ei} and references therein). Our main
original contribution is the new method for an accurate determination of the
density of states over the orders of magnitude requested by the problem
for a statistically meaningful evaluation of the integral.

In order to test our proposal, we have studied the $\mathbb{Z}(3)$
spin model at finite density~\cite{Karsch:1985cb,Kim:2005ck}, in which the
sign problem disappears in a dual reformulation. The existence of a
dual representation that is free from the sign problem enables one to
devise an exact algorithm~\cite{Mercado:2012yf} that can be used to
cross-check results obtained with our method. The results presented
below extend our original work~\cite{Langfeld:2014nta}, on which this
contribution is mostly based.

\section{The $\mathbb{Z}(3)$ spin model}
At strong coupling and for large fermion mass, for finite temperature
and non-zero chemical potential QCD is described by the three-dimensional
spin model 
\begin{eqnarray}
Z(\mu) = \sum _{\{\phi\}} \; \exp \Bigl\{ \tau \sum _{x,\nu } \left(
  \phi_x \, \phi^\ast _{x+\nu} + \mathrm{c.c.} \right) + \sum_x \, \Bigl( \eta \phi_x + \bar{\eta } \phi^\ast _x
\Bigr) \Bigr\} 
= \sum _{\{\phi\}} \;\exp \Bigl\{ S_s[\phi] + S_h[\phi] \Bigr\}  \ ,
\end{eqnarray}
with $\phi \in \mathbb{Z}(3)$ the spin variable. The spin
interaction is nearest-neighbour and is weighted by the coupling
$\tau$. The couplings $\eta = \kappa e^{\mu}$ and $\bar{\eta} = \kappa
e^{- \mu}$ are related to the chemical potential $\mu$ and to the
fermion hopping parameter $\kappa$. In the expression defining the
partition function, $\mathrm{c.c.}$ indicates the complex conjugate of
the spin-spin interaction. $S_s$ and $S_h$ are respectively the
spin-spin interaction and the contribution weighted by $\{\eta,\bar{\eta}\}$.

At $\kappa = 0$, the model reduces to the three-state Potts model,
which is known to have a first order phase transition at $\tau_c \simeq
0.18$. For $\tau < \tau_c$ the system is in the $\mathbb{Z}(3)$
symmetric phase, while for $\tau > \tau_c$ the system is in a broken
symmetry phase. At $\mu = 0$, the transition persists for a small non-zero
$\kappa$, to terminate in a tricritical point. More details on the
phase structure are provided in~\cite{Mercado:2011ua} and references therein.

At non-zero $\mu$, the action $S = S_s + S_h$ is complex. In fact,
while $S_s^{\ast} = S_s$, $S_h^{\ast}[\mu] = S_h[-\mu] \ne S_h[\mu]$
for any $\mu \ne 0$. However,
because of the symmetry $\phi \leftrightarrow \phi^{\ast}$,  the
partition function is real. This can be seen explicitly if we
reformulate the problem in terms of the number of spins that are
equal to each of the three cubic roots of unity $\{1,\ z=e^{i 2\pi/3}, \
z^{\ast}=e^{-i 2\pi/3}\}$. Introducing the variables 
\begin{eqnarray}
N_0 = \sum_x \delta \Bigl(\phi(x),1 \Bigr) \ , \
N_+=\sum_x \delta \Bigl(\phi(x),z \Bigr) \ , \
N_-= \sum_x \delta \Bigl(\phi(x),z^\ast
\Bigr) \ ,
\end{eqnarray}
which fulfil the constraint $N_0 + N_+ + N_- = N$ ($N$ being
the number of sites), $S_h$ can be written as
\begin{eqnarray}
S_h = \kappa \left[ \left( 2 \, N_0 - N_+ - N_- \right) \,
  \hbox{cosh}(\mu) \right.
 + \left. i \sqrt{3} \, (N_+ - N_-) \,
  \hbox{sinh}(\mu) \right] \; 
\end{eqnarray}
and $Z(\mu)$ takes the form
\begin{eqnarray}
\label{eq:zmudn}  
 Z(\mu) &=& \sum _{\{\phi\} } \exp \Bigl\{ S[\phi]  \; + \; 
 \kappa  \left( 3N_0 - V \right) \, \hbox{cosh}(\mu)
 \Bigr\} \cos \Bigl( \sqrt{3} \, \kappa \, \Delta N \,  \hbox{sinh}(\mu)
 \Bigr) \; ,
\end{eqnarray}
with $\Delta N = N_+ - N_-$ being the difference of the spins
aligned along the roots with positive and negative imaginary
part. Although real, $Z$ contains an oscillating term, which is at the
origin of the sign problem in this approach to the model.

The $\mathbb{Z}(3)$ model admits a reformulation in terms of
which there is no sign problem. This dual model has been used to
construct an algorithm that allows to simulate the system without
incurring in cancellations. The existence of an algorithm that
can be trusted in the high-$\mu$ regime makes the $\mathbb{Z}(3)$
model an ideal testbed for alternative approaches.

\section{Computing the density of states}
If we define a generalised density of state $\rho(n)$ as
\begin{eqnarray}
\rho (n)  = \sum _{\{\phi\}} 
\delta \Bigl( n, \Delta N[\phi] \Bigr) 
\exp \Bigl\{ S[\phi]  
+  \kappa  \Bigl( 3N_0 [\phi] - V \Bigr) \, \hbox{cosh}(\mu)
 \Bigr\} \ ,
\end{eqnarray}
the partition function $Z(\mu)$ takes the form
\begin{eqnarray}
Z(\mu)  =  \sum_{ n} \; \rho ( n)  \cos \Bigl( \sqrt{3} \, \kappa
\,   \hbox{sinh}(\mu)  n \Bigr) \ .
\end{eqnarray}
$\rho(n)$ being always positive, with this definition we have isolated the oscillating contribution
from the non-oscillating one. Hence, if we are able to determine
$\rho(n)$ with sufficiently high accuracy, we can in principle try to do
the oscillating sum. The needed accuracy is set by the simulation
parameters. However, the stronger the oscillation, the higher is the
accuracy requested to overcome the noise coming from cancellations of
positive and negative contributions. 

The Local Linear Relaxation (LLR) algorithm~\cite{Langfeld:2012ah} (see~\cite{Pellegrini:2014dva}
for recent developments) has been proven to give an accurate
determination for the density of states in gauge theories and in spin
systems~\cite{Guagnelli:2012dk}. The method is based on a linear
approximation of the logarithm of the density of states in a sufficiently small
interval of variation of the independent variable, with the angular coefficient determined with a recursive
relation. The full density of states can be reconstructed by imposing
continuity of the piecewise approximations at the edges of the
intervals. 

Following~\cite{Langfeld:2014nta},  we determine the generalised density of states 
$\rho(n)$ by using a modification of this algorithm.
Since $\phi \leftrightarrow \phi^{\ast}$ implies $\rho(-n) = \rho(n)$, we
only need to determine $\rho(n)$ for $n \ge 0$. For obtaining this quantity,
we formulate the ansatz  
\begin{equation}
\rho (n) \; = \; \prod_{i=0}^n \exp \{ - a_i \}  
\end{equation}
and use the LLR algorithm to determine the $a_n$. The procedure goes
as follows. We define $n$-restricted expectation values of a
function $F(a_n)$ as
\begin{eqnarray} 
\lb F \rb (a_n) = \frac{1}{\cal N} \sum _{\{\phi\}} \; F\left( 
\Delta N\left[\phi  \right]  \right) \; \theta  (\Delta N,n)\; \exp \{ a_n \}
\exp \Bigl\{ S[\phi]  
\; + \; \kappa  \Bigl( 3N_0 [\phi] - V \Bigr) \, \hbox{cosh}(\mu)
 \Bigr\} \; , 
\end{eqnarray}
where $\theta (\Delta N,n) = 1 $ for $ \vert \Delta N[\phi] - n \vert \le 1 $
and $\theta (\Delta N,n)  = 0 $ otherwise. ${\cal N}$ is a
normalisation factor such that $\lb 1 \rb = 1$. 
The double bracket expectation values can be computed using standard Monte-Carlo
methods. In particular, we can use $\lb \Delta N \rb$ and $\lb
\Delta N^2 \rb$ to obtain the $a_n$ in each interval $[n - 1; n + 1]$
starting from a trial $a_n^0$. This can be achieved using the
Newton-Raphson recursion
\begin{eqnarray} 
a_n^{k+1} \; = \; a_n^k \; - \; \frac{ \lb \Delta N \rb (a_n^k) }{ 
\lb \Delta N^2 \rb (a_n^k)} 
\end{eqnarray}
that has been adapted from~\cite{Langfeld:2012ah}. In this way,
$\rho(n)$ is determined up to the free parameter $a_0$, which can be
fixed by imposing the normalisation condition $\rho(0) = 1$.

Since a Monte-Carlo averaging is involved in the recursion, the $a_n$
will be determined up to a statistical error. In order to account for
this error in a realistic way, a bootstrap procedure has been used
employing around 100 independent determinations of $a_n$ for each $n$. 

\begin{figure}
\begin{tabular}{cc}
\includegraphics[width=.45\textwidth]{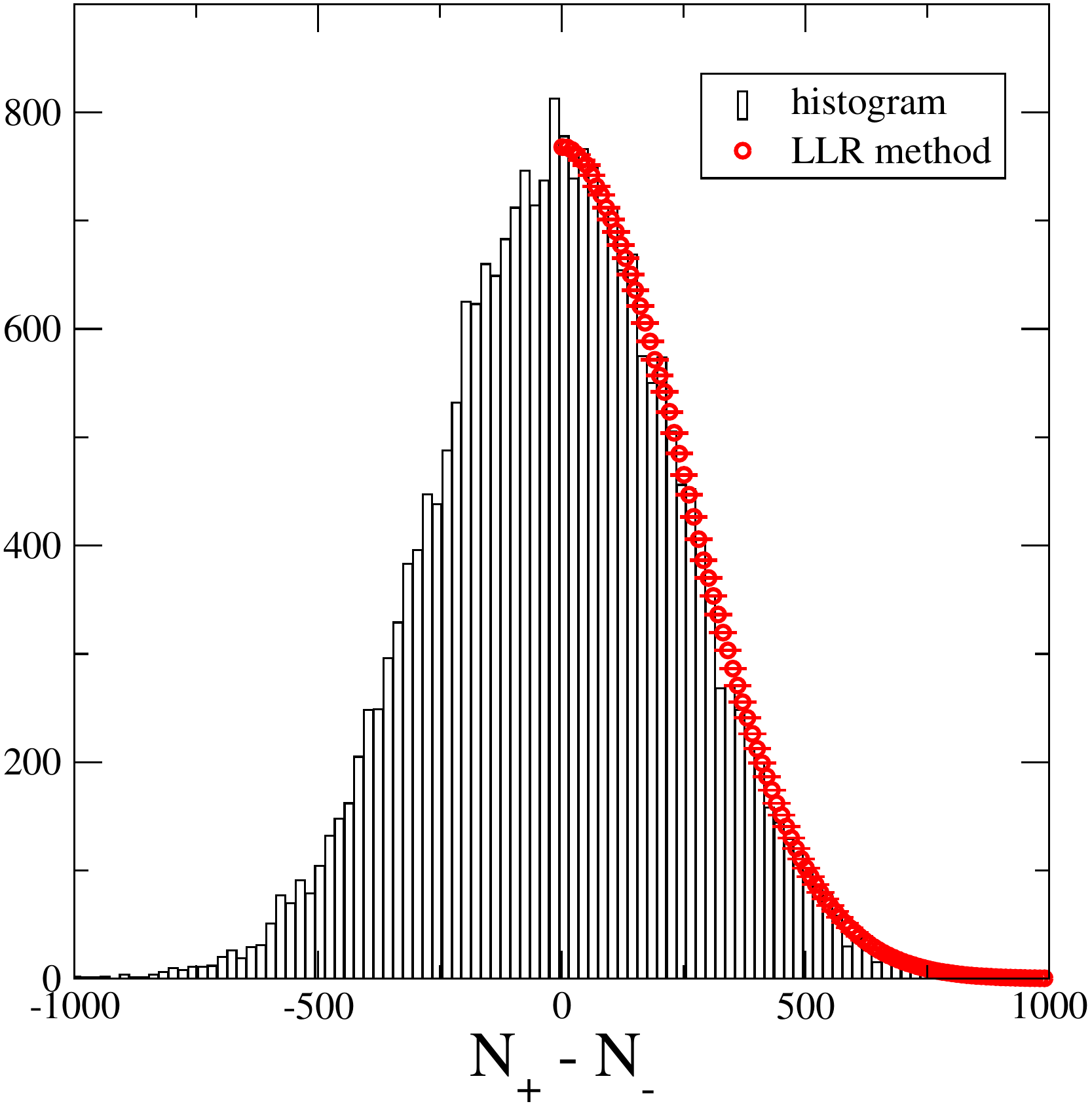}~~~~~~~~~~~~~~& 
\includegraphics[width=.45\textwidth]{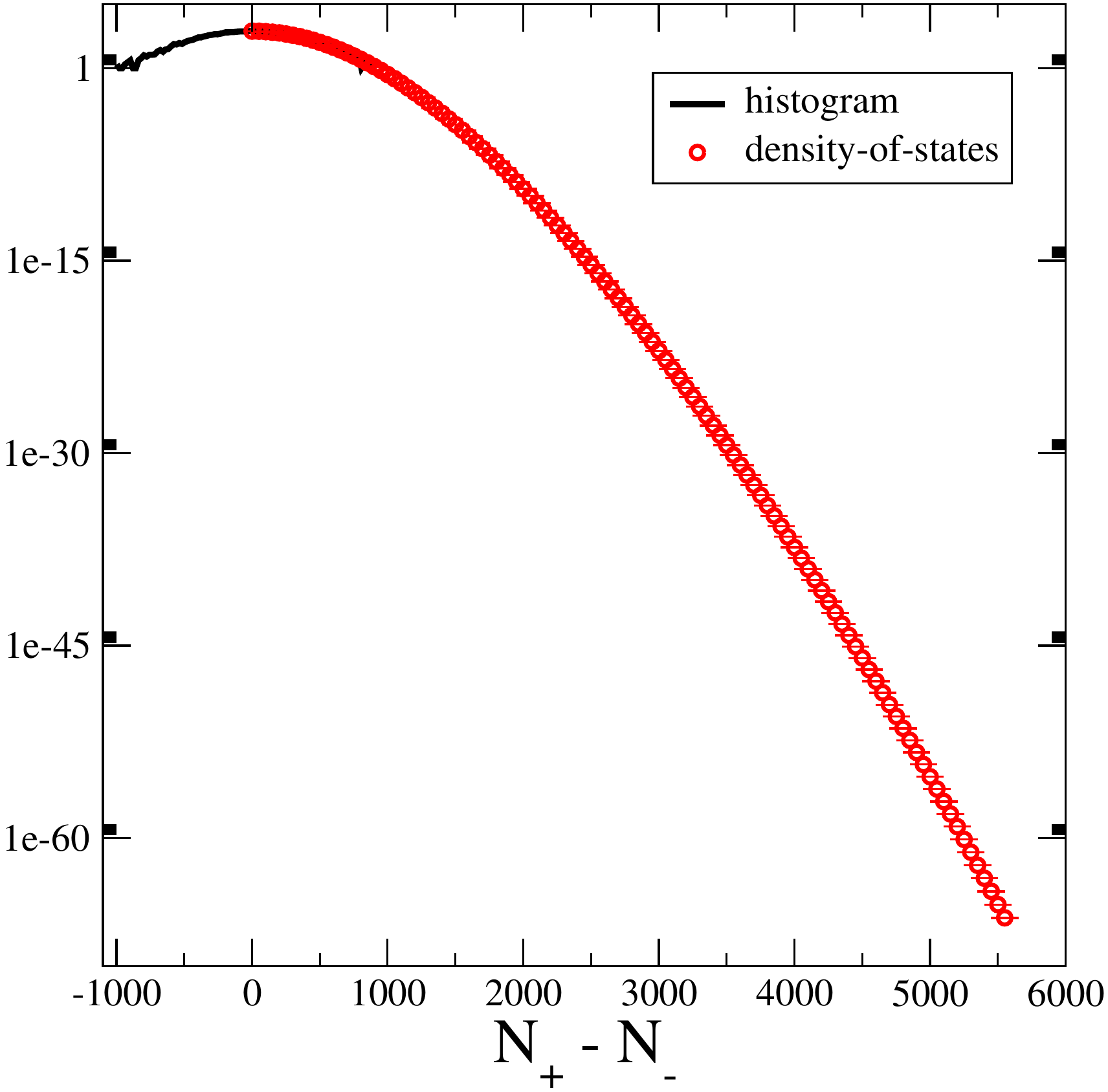}\\
{\Large (a)} & {\Large (b)}
\end{tabular}
\caption{(a) Comparison of the density of states reconstructed from a
  calculation with the worm algorithm and determined directly with
  the LLR method; (b) The full extension of the determination of the
  density of states with the LLR method. Both plots are for $\tau =
  0.17$ and $\kappa = 0.05$ at $\mu = 0$ on a $24^3$ lattice.\label{fig:1}}
\end{figure}

Fig.~\ref{fig:1}~(a) provides an example of a typical determination of
the density of states. In order to check the correctness of the
calculation, we have performed a simulation with the worm
algorithm. The two methods agree in the whole range for which the
histogram of the density of states can be reliably extracted from the
latter simulation (up to $\Delta N \simeq 1000$). The LLR method
allows us to go well beyond this value (we stopped our determination
at $\Delta N = 5500$), obtaining a density of states that spans well
over 60 orders of magnitude (see Fig.~\ref{fig:1}~(b)).

\section{Performing oscillating sums}
In this section, we provide evidence that the density of states
determined with the LLR method has sufficient accuracy for performing
directly oscillating sums even in regions in which the sign problem is severe.
The severity of the sign problem is measured by the expectation value
of the phase factor $O(\mu)$. This is given by
\begin{equation}
O(\mu) \; = \; \frac{ 
\sum_{ n} \; \rho ( n) \; \cos \Bigl( \sqrt{3} \, \kappa 
\,   \hbox{sinh}(\mu) \; n \Bigr) }{ \sum_{ n} \; \rho ( n) } \; = \;
\frac{Z(\mu)}{Z(0)} \ .
\end{equation}
Values of $O(\mu)$ close to one mean that the sign problem is mild;
conversely, $O(\mu) \ll 1$ means that the system is afflicted by 
a severe sign problem. 

$O(\mu)$ can be computed using a worm update at various $\mu$ and a
snake algorithm~\cite{deForcrand:2000fi} to solve the overlap problem
that arises when taking the ratio of partition functions at
significantly different values of $\mu$. This calculation
is not afflicted by the sign problem. Within the LLR method, $O(\mu)$
can be computed directly using the numerical determination of $\rho$.
We have performed a simulation at $\tau = 0.1$ for $\mu \le 2$. We
have found that the approach of reconstructing $O(\mu)$ directly from
the numerical data for $\rho$ does not provide the required precision on the
final result to make it statistically different from zero when the
sign problem is significant. A better
technique uses a polynomial interpolation of the logarithm of the
density of states\footnote{A similar method has been used
  in~\cite{Azcoiti:2002vk}.}. More in details, we can write   
$\ln \rho (n) \; = \; \sum _{k=0}^p c_k \, n^{2k}$, 
where we have imposed the constraint that the logarithm of the density
of states is even for $n \to -n$. We have used interpolations up to
$2p = 8$, finding that the result is very stable and only $c_0$ and
$c_2$ are significantly different from zero (within errors). All fits
provide acceptable values of $\chi^2$/dof. Using this semi-analytical
procedure, we were able to obtain an agreement of the phase factor up
to the maximum simulated value $\mu = 2$, where $O(\mu) \simeq
10^{-16}$, indicating a strong sign problem. We refer
to~\cite{Langfeld:2014nta} for details.

\begin{figure}
\begin{tabular}{cc}
\includegraphics[width=.45\textwidth]{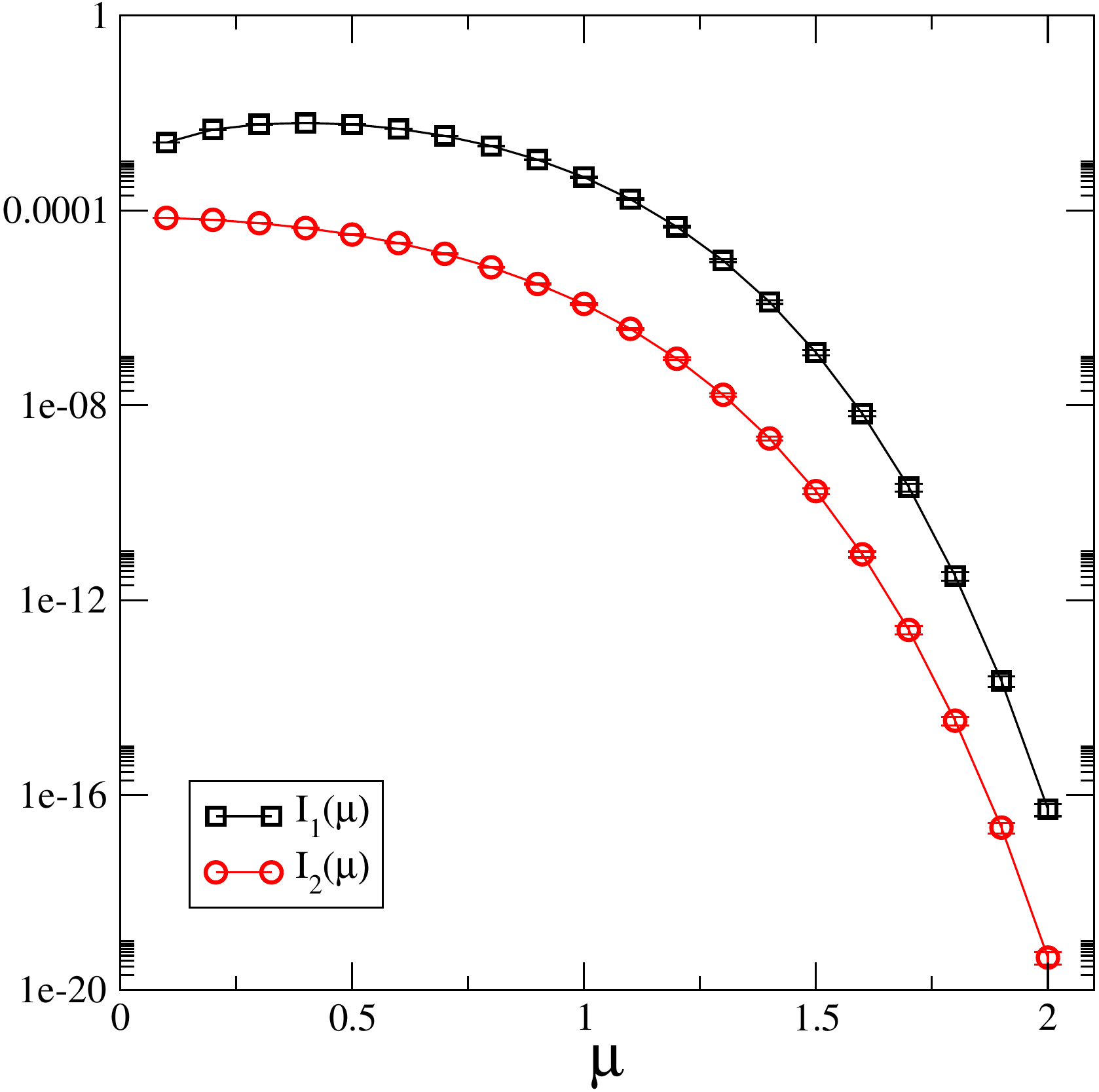}~~~~~~~~~~~~~~& 
\includegraphics[width=.45\textwidth]{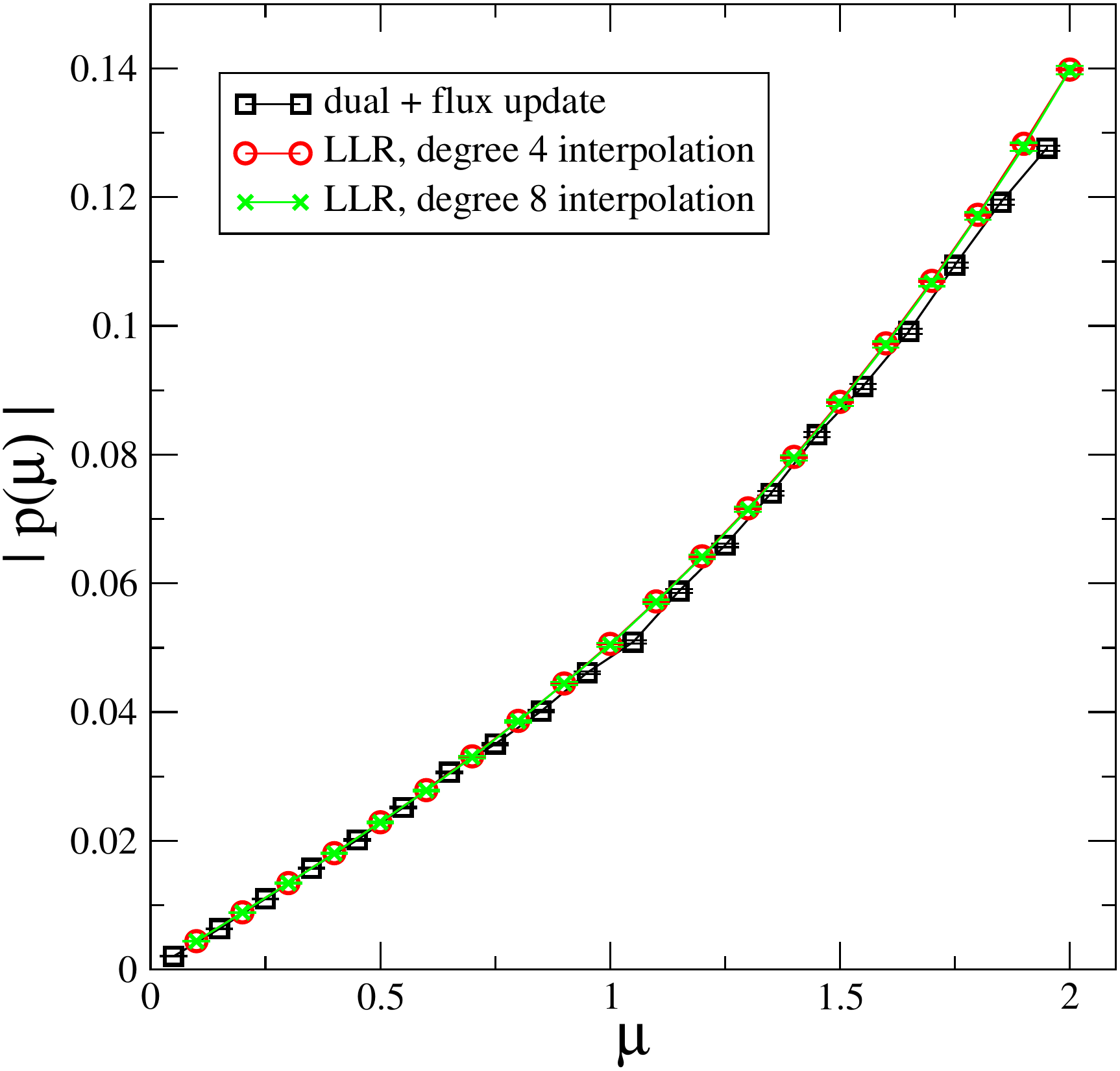}\\
{\Large (a)} & {\Large (b)}
\end{tabular}
\caption{(a) The numerator $I_1/Z(0)$ and the denominator $I_2/Z(0)$ of the
  phase twist $p(\mu)$; (b) The modulus of the phase twist. Both plots are for
  $\tau = 0.01$, $\kappa = 0.01$ (symmetric phase) on a $24^3$ lattice.\label{fig:2}}
\end{figure}
Another oscillating quantity is the phase twist
\begin{equation}
p(\mu ) \; = \; i \, \frac{ \sqrt{3} }{ V } \; \langle N_+ -
N_- \rangle \ ,
\end{equation}
which in our formalism can be expressed as
\begin{equation}
p(\mu ) = \frac{1}{V}\frac{\sum _n \rho(n) \; n \; \sin \Bigl( \kappa \sqrt{3} 
\, \sinh (\mu) \; n \Bigr)}
{\sum _n \rho(n) \;  \cos \Bigl( \kappa \sqrt{3} 
\, \sinh (\mu) \; n \Bigr) } = \frac{1}{V} \frac{I_1}{I_2} \ .
\end{equation}
Fig.~\ref{fig:2}~(a) shows a plot of the numerator $I_1$ and of the
denominator $I_2$ in the definition of $p(\mu)$, both normalised by
dividing them by $Z(0)$. These two quantities vary over several orders of
magnitude and following strong cancellations become
of order $10^{-16}-10^{-19}$ at $\mu = 2$. Hence, even if their final
ratio is of order $10^{-1}$, this is still a non-trivial test of how
cancellations are resolved by our method. Fig~\ref{fig:2}~(b) shows
good agreement between a semi-analytical determination with the
fitted density of states and a direct Monte-Carlo simulation using a
sign problem-free algorithm.

\section{Discussion and conclusions}
In this contribution, we have proposed a general method for simulating
systems with the sign problem. The method consists in a precise
determination of a generalised density of states using the LLR
algorithm and a semi-analytical calculation of oscillating
quantities. The method has been tested on the $\mathbb{Z}(3)$ spin model,
where it has been shown to reproduce the numerical results obtained
with a dual algorithm, which does not suffer from the sign
problem. 

The same model has been studied with similar techniques in another
contribution~\cite{Mercado:2014dva} on $10^3$ lattices for two
different sets of parameters corresponding to the deep symmetric phase
(which is the regime of our investigation) and to a situation in which
the system is close to the phase
transition. Ref.~\cite{Mercado:2014dva} confirms our conclusions deep  
in the symmetric phase. Near the phase transition, the authors observe
good agreement for the phase twist obtained with the density of states
and with the dual algorithm in a wide range of $\mu$, with some
deviations appearing at very high values of $\mu$. These deviations
could be due to the fact that the proposed ansatz for the density of
states might be not appropriate in that regime or to a loss
of efficiency of the dual algorithm. In any case, the discrepancy
needs to be understood (and resolved) by performing dedicated
simulations. 

\section*{Acknowledgements}
We thank V. Azcoiti, Ph. de Forcrand, C. Gattringer, J. Greensite, Y. Mercado,
R. Pellegrini, A. Rago and P. T\"orek for discussions. This work is supported by STFC under the DiRAC
framework. We are grateful for the support from the HPCC Plymouth,
where the numerical computations have been carried out. KL is supported by 
the  Leverhulme Trust (grant RPG-2014-118) and STFC (grant ST/L000350/1). BL is
supported by STFC (grant ST/G000506/1). 

\bibliographystyle{JHEP}
\bibliography{density_ym}

\end{document}